\documentclass[12pt]{article}
\usepackage[dvips]{graphicx}
\usepackage{textcomp}

\parskip=\medskipamount

\def\be{\begin{equation}}
\def\ee{\end{equation}}
\def\bea{\begin{eqnarray}}
\def\eea{\end{eqnarray}}
\def\bean{\begin{eqnarray*}}
\def\eean{\end{eqnarray*}}
\def\r#1{(\ref{#1})} 
\def\la#1{\label{#1}}                                     
\def\c#1{\cite{#1}}
\def\o{\omega}
\def\l{\lambda}

\title{HexaKdV}
\author{Jeremy Schiff \\
        Department of Mathematics,  Bar-Ilan University\\
        Ramat Gan 52900, Israel\\
        email: schiff@math.biu.ac.il}
\date{September 2002}

\begin{document}

\maketitle

\begin{abstract}
An analog of the lattice KdV equation of Nijhoff {\em et al.} 
is constructed on a hexagonal lattice. 
The resulting system of difference equations exhibits soliton solutions with 
interesting local structure: there is a 
nontrivial phase shift on moving between adjacent lattice sites, 
with the magnitude of the shift tending to zero in the continuum limit.
\end{abstract}

Bobenko, Hoffmann and Suris \c{BHS}
have recently proposed that some effort should be made
to understand integrable systems of difference equations on lattices other
than ${\bf Z}^2$. In addition to being of theoretical interest, it is 
possible that in certain cases 
this could have physical significance \c{MER,KMG}. 
This letter examines the analog of Nijhoff {\em et al.}'s lattice
KdV equation \c{NC,me} on a hexagonal lattice, which we call
``hexaKdV'' for short. The soliton solutions of hexaKdV are 
constructed. HexaKdV solitons undergo a 
nontrivial phase shift on moving between adjacent lattice sites, 
with the magnitude of the shift tending to zero in the continuum limit.

Nijhoff {\em et al.}'s lattice KdV equation involves a single field 
$b$ defined at the vertices of the standard lattice ${\bf Z}^2$. 
The equation gives a relation between the values of the field at the 
4 vertices of each fundamental lattice plaquette. Writing $b_1$ for 
$b_{n,m}$, $b_2$ for $b_{n+1,m}$, $b_3$ for $b_{n+1,m+1}$ and $b_4$ for 
$b_{n,m+1}$ (so $b_1,b_2,b_3,b_4$ are values of the field as we go around 
a fundamental plaquette), the equation takes the form \c{me}
\be
\frac{b_1-b_2-b_3+b_4}{h} + 
\frac{-b_2+b_3+b_4-b_1}{k} - b_1b_2 + b_2b_3 - b_3b_4 + b_4b_1 = 0 \ .
\ee
Two features of this equation are important in the sequel. 
The first is that the equation 
is unchanged on replacing $b_1$ by $b_2$, $b_2$ by $b_3$, $b_3$ by $b_4$, 
$b_4$ by $b_1$, $h$ by $k$ and $k$ by $-h$.  
This symmetry is associated with rotating the lattice through 90\textdegree. 
Indeed, the equation can be written in the evidently symmetric, and
rather more compact, form 
\be \left( I - R + R^2 - R^3 \right) 
   \left( \frac{b_1-b_2}{h} - b_1b_2 \right) = 0\ ,
\la{dkdv}
\ee
where $R$ is the rotation (or replacement) operator. 
The second important feature to note about standard  lattice KdV is that 
the number of variables and the number of equations 
is properly balanced. Each equation relates between 4 different 
values  of the field, but each value of the field appears in 4 different
equations (as each vertex of the lattice belongs to 4 fundamental 
plaquettes). Thus lattice KdV is a set of equations that can be expected to 
yield a solution, given suitable initial/boundary data.

The hexaKdV system which will shortly be 
constructed also displays a rotation symmetry.
Consistent with this, each equation in hexaKdV  relates the 
6 values of the field at the vertices of each fundamental hexagonal 
plaquette. But now a problem arises: Each vertex of the lattice 
evidently belongs to only 3 fundamental plaquettes, so it seems there are 
twice as many variables as equations (each equation calls 6 variables,
but each variable only appears in 3 equations). 
The simplest imaginable solution to this
problem is that there should be {\em two} equations associated with each plaquette. 
In fact  there is a system of four equations associated with each plaquette,
but they are degenerate and should be considered like two. 

The hexaKdV system will be constructed on
the hexagonal lattice with vertices 
\be 
\{ n_1 q+n_2 h\o+n_3 k\o^2 \ : \ n_1,n_2,n_3 \in {\bf Z} , \ 
                      n_1+n_2+n_3=0~{\rm or}~1 \ \}\ , 
\ee
where $\o=e^{2i\pi/3}$, and $q,h,k$ are arbitrary positive reals.
See figure 1. 
The vertex $n_1 q+n_2 h\o+n_3 k\o^2 $ will be referred to in the sequel
simply as ``the vertex $n_1,n_2,n_3$''.
There are 3 kinds of edges, parallel to $1,\o,\o^2$ respectively.
Following the ideas of \c{BHS} and \c{NC}, we look for a $GL(2)$-valued function
$\Psi_{n_1,n_2,n_3}(\l)$ defined on the vertices of the lattice and 
dependent on the spectral parameter $\lambda$, satisfying 
the following equations, one for each edge in the lattice:
\bea
\Psi_{n_1+1,n_2,n_3}(\l) &= & 
\pmatrix{ 1 - q b_{n_1+1,n_2,n_3} & q \cr 
      q\l + b_{n_1,n_2,n_3} - b_{n_1+1,n_2,n_3} 
      -q b_{n_1,n_2,n_3}b_{n_1+1,n_2,n_3}   &  1 + q b_{n_1,n_2,n_3} \cr}
\Psi_{n_1,n_2,n_3}(\l),\nonumber\\ 
\Psi_{n_1,n_2+1,n_3}(\l) &= & 
\pmatrix{ 1 - h b_{n_1,n_2+1,n_3} & h \cr 
      h\l + b_{n_1,n_2,n_3} - b_{n_1,n_2+1,n_3} 
     -hb_{n_1,n_2,n_3}b_{n_1+1,n_2,n_3}   &  1 + h b_{n_1,n_2,n_3} \cr}
\Psi_{n_1,n_2,n_3}(\l) ,\\
\Psi_{n_1,n_2,n_3-1}(\l) &= &  
\pmatrix{ 1 - k b_{n_1,n_2,n_3-1} & k \cr 
      k\l + b_{n_1,n_2,n_3} - b_{n_1,n_2,n_3-1} 
      -k b_{n_1,n_2,n_3}b_{n_1,n_2,n_3-1}   &  1 + k b_{n_1,n_2,n_3} \cr}
\Psi_{n_1,n_2,n_3}(\l) .\nonumber
\eea
The consistency condition arising from the two different ways to go around
a fundamental plaquette ($n_1,n_2,n_3 \rightarrow n_1,n_2+1,n_3 
\rightarrow n_1,n_2+1,n_3-1 
\rightarrow n_1+1,n_2+1,n_3-1$ or  
$n_1,n_2,n_3 \rightarrow n_1+1,n_2,n_3 \rightarrow n_1+1,n_2,n_3-1 
\rightarrow n_1+1,n_2+1,n_3-1$, see figure 2) can be written 
$$
\pmatrix{1 - qb_4 & q \cr q\l + b_3-b_4-qb_3b_4 & 1 + qb_3 \cr}
\pmatrix{1 - kb_3 & k \cr k\l + b_2-b_3-kb_2b_3 & 1 + kb_2 \cr}
\pmatrix{1 - hb_2 & h \cr h\l + b_1-b_2-hb_1b_2 & 1 + hb_1 \cr}
$$
\be = \la{cons}\ee
$$
\pmatrix{1 - hb_4 & h \cr h\l + b_5-b_4-hb_5b_4 & 1 + hb_5 \cr}
\pmatrix{1 - kb_5 & k \cr k\l + b_6-b_5-kb_6b_5 & 1 + kb_6 \cr}
\pmatrix{1 - qb_6 & q \cr q\l + b_1-b_6-qb_1b_6 & 1 + qb_1 \cr}
$$
where here for brevity  $b_1$ stands for $b_{n_1,n_2,n_3}$, 
$b_2$ for $b_{n_1,n_2+1,n_3}$, $b_3$ for $b_{n_1,n_2+1,n_3-1}$,  
$b_4$ for $b_{n_1+1,n_2+1,n_3-1}$,  $b_5$ for $b_{n_1+1,n_2,n_3-1}$ and
$b_6$ for $b_{n_1+1,n_2,n_3}$, these being the 6 values of $b$ 
at vertices of a fundamental plaquette.  

\begin{figure}
\centerline{\includegraphics[width=80mm,angle=270]{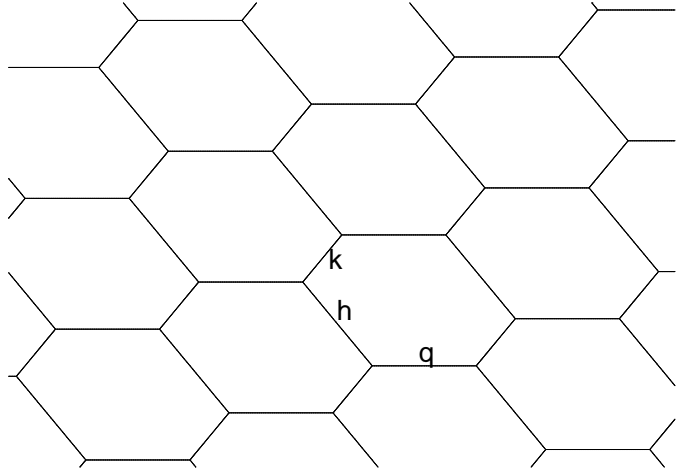}}
\caption{The hexagonal lattice}
\end{figure}

\begin{figure}
\vskip.2in
\centerline{\includegraphics[width=80mm]{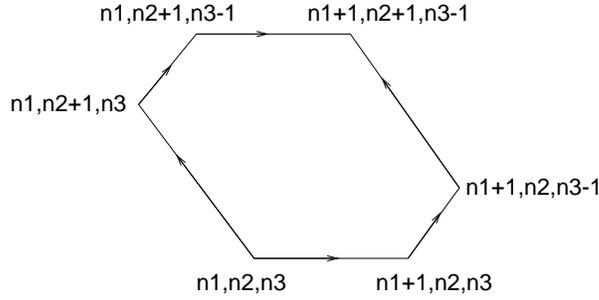}}
\caption{Two ways around a fundamental plaquette}
\end{figure}

A symbolic manipulator was used to multiply out \r{cons}. Four equations 
emerge. Since full rotational symmetry has been broken by the choice of a 
starting and ending point in the fundamental plaquette, the equations 
are not rotation invariant, but this can be rectified by taking 
suitable linear combinations and just the final symmetric form of the 
equations will be presented here. Writing  $R$ for the 60\textdegree rotation operator,
implemented by replacing $b_1$ by $b_2$, $b_2$ by $b_3$, $b_3$ by $b_4$, 
$b_4$ by $b_5$, $b_5$ by $b_6$, 
$b_6$ by $b_1$, $h$ by $k$ and $k$ by $q$ and $q$ by $-h$, the 
equations are:
\be 
   \left( I - R + R^2 - R^3 + R^4 - R^5 \right) 
   \left( \frac{b_1-b_2}{h} - b_1b_2 \right) = 0\ ,
\la{hkdv1}
\ee
\be 
   \left( I + R + R^2 + R^3 + R^4 + R^5 \right) 
   \left(\frac{(b_1-b_2)^2}{h} - b_1b_2(b_1-b_2) \right)
    = 0\ ,
\la{hkdv2}
\ee
\bea 
   \left( I - R + R^2 - R^3 + R^4 - R^5 \right) 
   \left( 
   \frac{2(b_1+b_2)}{qk} -b_1b_4(b_2+b_3) 
   \right. &&  \la{hkdv3}\\
   \left.
   +\frac{b_1b_3+b_1b_4+b_2b_3-b_2b_5-b_3b_4-b_3b_5}{h}
   \right)   
   &=& 0\ ,\nonumber
\eea
\bea 
   \left( I - R + R^2 - R^3 + R^4 - R^5 \right) 
   \left(
    \frac{2(b_1+b_2)^2}{qk}
   +b_1b_2(b_1^2+b_2^2-b_1b_2-2b_3b_4-b_4b_1-b_2b_5)
   \right. && \la{hkdv4}\\
\left.
-\frac{b_1^3-b_2^3
+b_2^2(2b_1+b_5)   - b_1^2(b_4+2b_2)   +2b_3^2(b_4-b_2)
+2(b_1b_2b_6+b_2b_3b_5-b_1b_2b_3-b_1b_3b_4)
}{h}
\right) &=& 0\ .
\nonumber\eea
Note the similarity of the first equation, \r{hkdv1}, with the 
standard lattice KdV equation \r{dkdv}. An analysis of the 
hexaKdV system \r{hkdv1}-\r{hkdv4} is quite straightforward with a 
symbolic manipulator. The first equation, \r{hkdv1}, is linear
in $b_1$, and can be used to eliminate $b_1$ provided its 
coefficient, $b_6-b_2+\frac1{h}-\frac1{q}$, does not vanish.
Proceeding along these lines it can be shown that
the solution set consists 
of two components, characterized by whether or not
$b_6-b_2+\frac1{h}-\frac1{q}$ vanishes.
One component is a dimension 3 set of solutions given 
by solutions of the 3 linear equations 
\be
\left.
\begin{array}{rcl}
   b_6 - b_2 + \frac1{h} - \frac1{q} &=& 0 \\
   b_2 - b_4 + \frac1{q} + \frac1{k} &=& 0 \\
 \left(  \frac1{h^2} - \frac1{q^2} \right) b_1
+\left(  \frac1{q^2} - \frac1{k^2} \right) b_3
+\left(  \frac1{k^2} - \frac1{h^2} \right) b_5 
&=& \left( \frac1{q} - \frac1{h}  \right)
    \left( \frac1{k} + \frac1{q}  \right)
    \left( \frac1{h} + \frac1{k}  \right)
\end{array}
\right\}\ .
\la{3dset}
\ee
The other component is
a dimension 4 set of solutions in which 
$b_1$ is determined by 
\be
b_1= 
\frac 
{{b_{3}}\,{b_{4}} - {b_{3}}\,{b_{2}} - {b_{5}}\,{b_{4}} + {b_{5}}\,{b_{6}} 
+  \frac {{b_{5}} - {b_{4}} + {b_{2}}}{h}  
+  \frac { - {b_{3}} + {b_{4}} - {b_{6}}}{q}  
+  \frac { - {b_{3}} + {b_{5}} - {b_{6}} + {b_{2}}}{k} }
{{b_{6}} - {b_{2}} + \frac {1}{h}  - \frac {1}{q}  }\ , 
\ee
and $b_2,b_3,b_4,b_5,b_6$ satisfy the single constraint 
\bea
0&=&  {b_{2}}\,{b_{5}}\,{b_{6}} + {b_{6}}\,{b_{3}}\,{b_{4}} 
- {b_{2}}\,{b_{5}}\,{b_{4}} - {b_{6}}\,{b_{3}}\,{b_{2}} 
+ {b_{5}}\,{b_{4}}^{2} - {b_{3}}\,{b_{4}}^{2} 
+ {b_{3}}\,{b_{4}}\,{b_{2}} - {b_{5}}\,{b_{4}}\,{b_{6}}  \nonumber\\
&&+ \frac 
{{b_{4}}\,{b_{6}} - {b_{4}}^{2} - {b_{6}}\,{b_{2}} + {b_{5}}\,{b_{6}} 
- {b_{3}}\,{b_{6}} + {b_{4}}\,{b_{2}} + {b_{3}}\,{b_{4}} - {b_{5}}\,{b_{4}}}{q}
+ \frac {{b_{6}} + {b_{5}} - {b_{3}} - 2\,{b_{4}} + {b_{2}}}{h\,k}  
\nonumber\\
&&+ \frac 
{ - {b_{3}}\,{b_{6}} + {b_{5}}\,{b_{6}} - 2\,{b_{5}}\,{b_{4}} + 
2\,{b_{3}}\,{b_{4}} - {b_{3}}\,{b_{2}} + {b_{2}}\,{b_{5}}}{k} 
+ \frac { - {b_{2}} - {b_{3}} + {b_{6}} + {b_{5}}}{h\,q}  
  \\
&&+ \frac 
{{b_{3}}\,{b_{4}} + {b_{2}}\,{b_{5}} + {b_{4}}^{2} - {b_{5}}\,{b_{4}} - 
{b_{4}}\,{b_{2}} - {b_{3}}\,{b_{2}} - {b_{4}}\,{b_{6}} + {b_{6}}\,{b_{2}}}{h} 
+ \frac { - {b_{6}} + 2\,{b_{4}} - {b_{2}} - {b_{3}} + {b_{5}}}{q\,k}  
\nonumber\\
&& + \frac { - {b_{3}} + {b_{5}} - {b_{6}} + {b_{2}}}{k^{2}}  \ .
\nonumber
\eea
It can be checked that the 4 dimensional set of solutions includes the 
3 dimensional set obtained by rotating the solution of \r{3dset}, 
i.e. the solutions of
\be
\left.
\begin{array}{rcl}
   b_1 - b_3 + \frac1{k} + \frac1{h} &=& 0\\
   b_3 - b_5 - \frac1{h} + \frac1{q} &=& 0\\ 
 \left(  \frac1{k^2} - \frac1{h^2} \right) b_2
+\left(  \frac1{h^2} - \frac1{q^2} \right) b_4
+\left(  \frac1{q^2} - \frac1{k^2} \right) b_6 
&=& \left( \frac1{h} + \frac1{k}  \right)
    \left( - \frac1{q} + \frac1{h}  \right)
    \left( \frac1{k} + \frac1{q}  \right)
\end{array}
\right\}\ .
\la{3dset2}
\ee

Thus, to summarize, we have arrived at the hexaKdV system. This consists of 
the system \r{hkdv1}-\r{hkdv4} of 4 polynomial equations in 6 variables
on each plaquette of the lattice. Because 
the system \r{hkdv1}-\r{hkdv4} is degenerate, 
in the sense that it has a 4 dimensional solution set, it is more 
appropriate to think of hexaKdV as specifying 2 constraints for each plaquette.
Thus, as explained above, 
there is a proper balance between the number of variables and the 
number of constraints in hexaKdV.

\vskip.2in

\noindent{\em Soliton solutions.} A direct computation shows that 
\r{hkdv1}-\r{hkdv4} has the following 
two parameter family of solutions:
\be
\begin{array}{rcl}
b_1 &=& C \tanh\left(z\right)\ ,  \\
b_2 &=& C \tanh\left(z+\tanh^{-1}(hC)\right)\ ,  \\
b_3 &=& C \tanh\left(z+\tanh^{-1}(hC)+\tanh^{-1}(kC)\right)\ ,  \\
b_4 &=& C \tanh\left(z+\tanh^{-1}(hC)+\tanh^{-1}(kC)
             +\tanh^{-1}(qC)\right)\ ,  \\
b_5 &=& C \tanh\left(z+\tanh^{-1}(kC)+\tanh^{-1}(qC)\right)\ ,  \\
b_6 &=& C \tanh\left(z+\tanh^{-1}(qC)\right)\ .
\end{array}
\ee
Here $C,z$ are arbitrary. Such solutions on an individual plaquette
can be pasted together to give a full solution of hexaKdV
of form
\be
b_{n_1,n_2,n_3} =  C \tanh\left(
n_1 \tanh^{-1}(qC) + 
n_2 \tanh^{-1}(hC) -
n_3 \tanh^{-1}(kC) + 
z\right)\ ,
\la{sol}\ee
where again $C,z$ are arbitrary constants. To get some understanding into
the nature of the soliton solution, it is necessary to 
write it as a function 
of the standard Cartesian coordinates $x$ and $y$ of the vertex $n_1,n_2,n_3$.
These are given by
\be
x = n_1 q -\frac12(n_2 h+ n_3 k) \ , \qquad 
y = \frac{\sqrt{3}}{2} (n_2 h - n_3 k) \ .
\ee 
Writing $n_1,n_2,n_3$ in terms of $x,y$ and the quantity 
\be s=n_1+n_2+n_3 \ee
(which is $0$  or $1$),  the soliton becomes 
\be
b(x,y,s) = C\tanh(D(x+cy)+Es+z)  
\ ,
\la{sol2}\ee
where 
\bea
D &=& \frac{(k+h)\tanh^{-1}(qC)+h\tanh^{-1}(kC)-k\tanh^{-1}{hC}}
{qk+qh+hk}\ , \la{wa}\\
c &=& \frac{  (h-k)\tanh^{-1}(qC)+(h+2q)\tanh^{-1}(kC)+(k+2q)\tanh^{-1}{hC}   } 
{\sqrt{3}\left((k+h)\tanh^{-1}(qC)+h\tanh^{-1}(kC)-k\tanh^{-1}{hC}\right)} \ ,  
\la{sa}\\
E &=& \frac
{kh\tanh^{-1}(qC)-qh\tanh^{-1}(kC)+kq\tanh^{-1}{hC} } {qk+qh+hk} \ .
\eea
Before discussing this result, note that for sufficiently small $q,h,k$
(for which  $\tanh^{-1}(qC)\approx qC$ etc)
\bea
D &\approx& \frac{q(k+h)}{qk+qh+hk}\ , \la{lim1}\\
c &\approx& \frac{qk+3qh+2kh}{\sqrt{3}q(k+h)}\ ,\la{lim2}\\
E &\approx& \frac{qkh} {qk+qh+hk} \la{lim3}\ .
\eea
In the form \r{sol2}, the meaning of the soliton solution is quite clear. 
$c$ is the speed of the soliton, $C$ its amplitude, and equation \r{sa}
expresses the speed-amplitude relation. The dependence of $D$ on $C$, through
\r{wa}, shows that the width of the soliton also depends on the speed (or 
amplitude), a common phenomenon seen, for example, in the KdV equation.
The novel feature of hexaKdV solitons is, however, the dependence of
\r{sol2}  on $s$. This means that {\em there is a phase shift between the 
soliton solution on ``even'' ($s=0$) and ``odd'' ($s=1$) sites of the lattice}.
This interesting phenomenon distinguishes hexaKdV from  standard 
lattice KdV. Equation \r{lim3} shows that as the continuum limit is approached
($q,h,k$ tending to $0$), the dependence on $s$ becomes very weak, and 
indeed vanishes in the continuum. Equations \r{lim1} and \r{lim2} show that 
as $q,h,k$ tend to $0$ with constant ratios, the speed and width of the solitons 
tend to constant values, independent of amplitude. 
 
\vskip.2in

\noindent{\em The continuum limit.} Is hexaKdV in any sense a discretization of 
a PDE? Consider making the following replacements in \r{hkdv1}-\r{hkdv4}:
\be
\begin{array}{rcl}
b_1 &\rightarrow& b\left(x-\frac12q+\frac14 h -\frac14 k, 
                         y-\frac{\sqrt{3}}{4}h-\frac{\sqrt{3}}{4}k \right)\ ,  \\
b_2 &\rightarrow& b\left(x-\frac12q-\frac14 h -\frac14 k, 
                         y+\frac{\sqrt{3}}{4}h-\frac{\sqrt{3}}{4}k \right)\ ,  \\
b_3 &\rightarrow& b\left(x-\frac12q-\frac14 h +\frac14 k, 
                         y+\frac{\sqrt{3}}{4}h+\frac{\sqrt{3}}{4}k \right)\ ,  \\
b_4 &\rightarrow& b\left(x+\frac12q-\frac14 h +\frac14 k, 
                         y+\frac{\sqrt{3}}{4}h+\frac{\sqrt{3}}{4}k \right)\ ,  \\
b_5 &\rightarrow& b\left(x+\frac12q+\frac14 h +\frac14 k, 
                         y-\frac{\sqrt{3}}{4}h+\frac{\sqrt{3}}{4}k \right)\ ,  \\
b_6 &\rightarrow& b\left(x+\frac12q+\frac14 h -\frac14 k, 
                         y-\frac{\sqrt{3}}{4}h-\frac{\sqrt{3}}{4}k \right)\ . 
\end{array}
\la{repls}
\ee
Expanding the 4 resulting equations in a Taylor series in $q,h,k$ and retaining only 
leading orders (assuming $h,q,k$ all to be of the same order) gives only 2 distinct
equations. Equations \r{hkdv3} and \r{hkdv4} both give 
\be
b_y = \frac{qk+3qh+2kh}{\sqrt{3}q(k+h)}b_x\ .
\la{weq}\ee
This is consistent with the results on soliton solutions, as for small $q,h,k$ 
the soliton speed is given by $c$ as in \r{lim2}, and this is precisely
the factor that has just appeared in \r{weq}.  However, both \r{hkdv1}
and \r{hkdv2} reduce to another PDE, which is only consistent with 
\r{weq} if $q=h$. So far we have no understanding of why it should 
be necessary to impose
such a constraint for a consistent continuous limit, or whether there is 
maybe some  reason to ignore it. 

\vskip.2in

\noindent{\em Dual Discretizations of PDEs.}
Although it is a digression from the main topic, since it has been shown 
that hexaKdV is a discretization of \r{weq}, at least in the case $h=q$, we briefly
raise the question of how {\em ab initio} one might go about discretizing 
this (or another) PDE on
a hexagonal lattice, with the intention of getting equations relating the 6 values 
of the field around each hexagonal plaquette. 

The standard finite difference approach for discretizing a PDE involves 
writing a discrete equation to approximate the PDE at each vertex of the lattice
being used. The discrete equation involves the values of the functions appearing
in the PDE at the relevant vertex, as well as the values at neighboring vertices. 
The balance between the number of equations obtained and the number of variables is 
automatic, as both equal the number of lattice vertices in the relevant domain.

Here we follow a different ``dual'' approach. The aim is to approximate the PDE on a 
fundamental lattice plaquette (or, more precisely, at the center of gravity of a
lattice plaquette), using the values of the function on lattice vertices. In the
case of a hexagonal lattice, this means it is necessary to approximate derivatives of 
$b$ at the point $(x,y)$ using the 6 values of $b$ that appeared in \r{repls}. 
This is straightforward. For example it can be shown  that under the 
replacements \r{repls}
\bea
\frac{b_4-b_1+b_6-b_3}{2q}
&=&  
b_x(x,y) 
+O(q^2,h^2,k^2,hk,qk,qh)\ , \la{fd1}\\  
\frac1{2\sqrt{3}}\left(
\frac{b_5-b_2-b_6+b_3}{k} + \frac{b_4-b_1-b_5+b_2}{h}
\right)
&=&  
b_y(x,y)
+O(q^2,h^2,k^2,hk,qk,qh)\ .\la{fd2}
\eea
Thus the PDE $b_y=Mb_x/\sqrt{3}$ can apparently be discretized by the difference
equation
\be
\frac{b_5-b_2-b_6+b_3}{k} + \frac{b_4-b_1-b_5+b_2}{h}
=
M \  \frac{b_4-b_1+b_6-b_3}{q}\ .
\la{pap1}\ee
However, there is now a counting problem of exactly the type mentioned before. 
The PDE has been replaced by a single difference equation for each plaquette, with 
the difference equation ``calling'' 6 values of the field. But, each value of 
the field only appears in 3 equations, so the number of variables is twice the 
number of equations. The resolution of this is that there is
a constraint that needs to be imposed.
Working to the same order as the approximations \r{fd1}-\r{fd2} it can be 
checked that  
\be
\left(\frac1{h}+\frac1{q}\right)(b_4-b_1) - 
\left(\frac1{k}+\frac1{h}\right)(b_5-b_2) +
\left(\frac1{q}+\frac1{k}\right)(b_6-b_3) = 0 +O(q^2,h^2,k^2,hk,qk,qh)\ .
\ee
Thus equation \r{pap1} on each plaquette must be supplemented by the 
constraint
\be
\left(\frac1{h}+\frac1{q}\right)(b_4-b_1) - 
\left(\frac1{k}+\frac1{h}\right)(b_5-b_2) +
\left(\frac1{q}+\frac1{k}\right)(b_6-b_3) = 0  \ .
\la{pap2}\ee
This resolves the counting problem. 

Dual discretizations of PDEs may well be appropriate in a variety of
settings. It is certainly no surprise that they arise in 
discretizations of equations in the KdV hierarchy, which are, in a natural
way, zero curvature equations. In conclusion of this section, note that
the constraint equation \r{pap2} looks very similar to what is obtained 
by applying all the necessary rotation operators just to the term 
$(b_1-b_2)/h$ in \r{hkdv1}. But the differences turn out to be significant;
for hexaKdV there does not seem to be a natural division of the equations into
a constraint part and a dynamic part, as there is with the simple discretization
comprised of \r{pap1} and \r{pap2}. 

\vskip.2in

\noindent{\em OctaKdV and other lattices.} 
The work presented in this letter can be extended to other lattices. We report
briefly on the extension to a lattice of alternating octahedra and rectangles
(with the octahedra having 4 pairs of parallel sides of equal length). On the 
rectangular plaquettes standard lattice KdV must hold. A simple counting
argument then shows that there should be 3 equations (in 8 variables) associated to 
each octagonal plaquette. Looking at the analog of \r{cons} with 4 matrices in 
each side gives a complicated system of 6 polynomial equations in 8 variables, 
with the first being an obvious extension of \r{dkdv} and \r{hkdv1}.
We hypothesize that this system has a 5 parameter solution set, and thus should
be thought of as just 3 constraints. As of yet we have not been able to verify this
in general, the memory requirements for the algebra being substantial. We have, 
however, looked at a number of reductions, in which one variable is assumed to 
vanish; in such reductions there are 4 parameter solution sets, which is
consistent with the hypothesis. 

It may be possible to obtain some general results on the algebraic 
system obtained from the generalization of \r{cons} with $n$ matrices on each 
side, relevant to a plaquette with $2n$ sides. Presumably soliton solutions on
more general lattices will also exhibit 
``small scale'' phase shifts as we move round a plaquette, as there are for
hexaKdV.  These phase shifts may well be indicative of lattice structure. 

It remains a mystery whether there is a way to formulate
integrable systems of difference equations 
on lattices with plaquettes with odd numbers of sides,
particularly triangular lattices. 

\section*{Acknowledgments}

This work was supported by the Israel National Science Foundation. 
I thank David Kessler for numerous discussions.


\begin{thebibliography}{9}

\bibitem{BHS}
A.I.Bobenko, T.H.Hoffmann and Yu.B.Suris, 
{\em Hexagonal Circle Patterns and Integrable Systems: 
Patterns with the Multi-Ratio Property and Lax Equations 
on the Regular Triangular Lattice},
{\sl Int.Math.Res.Not.} {\bf 2002} 111-164. 

\bibitem{MER}
J.L.Mar\'in, J.C.Eilbeck and F.M.Russell,
{\em Localized Moving Breathers in a 2D Hexagonal Lattice},
{\sl Phys.Lett.A} {\bf 248} (1998) 225-229.

\bibitem{KMG}
P.G.Kevrekidis, B.A.Malomed and Yu.B.Gaididei,
{\em Solitons in Triangular and Honeycomb Dynamical Lattices
with the Cubic Nonlinearity},
{\tt arXiv:nlin.PS/0205045}. 

\bibitem{NC}
F.Nijhoff and H.Capel, 
{\em The Discrete Korteweg-de Vries Equation},
{\sl Acta Appl.Math.} {\bf 39} (1995) 133-158.

\bibitem{me}
J.Schiff, 
{\em Loop Groups and Discrete KdV Equations},
{\tt arXiv:nlin.SI/0209040}. 

\end{thebibliography}
\end{document}